\documentclass{article}

\usepackage{amssymb,amsmath,eucal}
\usepackage[dvips]{lscape,graphicx}

\newcommand{\ct}{\cite}
\newcommand{\lb}{\label}

\newcommand{\bc}{\begin{center}}
\newcommand{\ec}{\end{center}}
\newcommand{\bd}{\begin{displaymath}}
\newcommand{\ed}{\end{displaymath}}
\newcommand{\be}{\begin{equation}}
\newcommand{\ee}{\end{equation}}
\newcommand{\ba}{\begin{array}}
\newcommand{\ea}{\end{array}}
\newcommand{\bt}{\begin{tabular}}
\newcommand{\et}{\end{tabular}}
\newcommand{\un}{\underline}
\newcommand{\ov}{\overline}
\newcommand{\bp}{\begin{picture}}
\newcommand{\ep}{\end{picture}}
\newcommand{\bfi}{\begin{figure}}
\newcommand{\efi}{\end{figure}}
\newcommand{\ds}{\displaystyle}

\def\fun#1#2{\lower3.6pt\vbox{\baselineskip0pt\lineskip.9pt
\ialign{$\mathsurround=0pt#1\hfil##\hfil$\crcr#2\crcr\sim\crcr}}}

\begin{document}

\vspace{1cm}

\title{\Large \bf Multiple Point Model and Phase Transition Couplings \\
in the Two-Loop Approximation of Dual Scalar Electrodynamics}
\author{{\large  {\un{Larisa V. Laperashvili}},}
{\large \,\, Dmitri A. Ryzhikh}\\
\it Institute of Theoretical and Experimental Physics,
B.Cheremushkinskaya 25,\\
\it Moscow 117 259, Russia\\
and\\
{\large Holger Bech Nielsen}\\
\it Niels Bohr Institute, Blegdamsvej 17-21,\\
\it Copenhagen 0, Denmark}

\date{}

\maketitle
\pagenumbering{arabic}

The simplest effective dynamics describing the
confinement mechanism in the pure gauge lattice U(1) theory
is the dual Abelian Higgs model of scalar monopoles [1-3].

In the previous papers [4-6] the calculations of the U(1)
phase transition (critical) coupling constant were connected with the
existence of artifact monopoles in the lattice gauge theory and also
in the Wilson loop action model \ct{6}.
In Ref.\ct{6} we (L.V.L. and H.B.N.) have put forward the speculations
of Refs.[4,5] suggesting that the modifications of the form of
the lattice action might not change too much the phase transition value of the
effective continuum coupling constant.
In \ct{6} the Wilson loop action was considered in the
approximation of circular loops of radii $R\ge a$. It was shown that the
phase transition coupling constant is indeed approximately independent
of the regularization method: ${\alpha}_{crit}\approx{0.204}$,
in correspondence with the Monte Carlo simulation result on lattice \ct{7}:
${\alpha}_{crit}\approx{0.20\pm 0.015}$.

But in Refs.[2,3] instead of using the lattice or Wilson loop
cut--off we have considered the Higgs Monopole Model (HMM) approximating
the lattice artifact monopoles as fundamental pointlike particles described
by the Higgs scalar field.

\section{The Coleman-Weinberg effective potential for the Higgs \\
monopole model}

The dual Abelian Higgs model of scalar monopoles (shortly HMM),
describing the dynamics of confinement in lattice
theories, was first suggested in Ref.\ct{1}, and considers the
following Lagrangian:
\begin{equation}
    L = - \frac{1}{4g^2} F_{\mu\nu}^2(B) + \frac{1}{2} |(\partial_{\mu} -
           iB_{\mu})\Phi|^2 - U(\Phi),\quad              \lb{5y}
{\mbox{where}}\quad
 U(\Phi) = \frac{1}{2}\mu^2 {|\Phi|}^2 + \frac{\lambda}{4}{|\Phi|}^4
\end{equation}
is the Higgs potential of scalar monopoles with magnetic charge $g$, and
$B_{\mu}$ is the dual gauge (photon) field interacting with the scalar
monopole field $\Phi$. In this model $\lambda$ is the self--interaction
constant of scalar fields, and the mass parameter $\mu^2$ is negative.
In Eq.(\ref{5y}) the complex scalar field $\Phi$ contains
the Higgs ($\phi$) and Goldstone ($\chi$) boson fields:
\begin{equation}
          \Phi = \phi + i\chi.             \lb{7y}
\end{equation}
The effective potential in the Higgs Scalar ElectroDynamics (HSED)
was first calculated by Coleman and Weinberg \ct{9} in the one--loop
approximation. The general method of its calculation is given in the
review \ct{10}. Using this method, we can construct the effective potential
for HMM. In this case the total field system of the gauge ($B_{\mu}$)
and magnetically charged ($\Phi$) fields is described by
the partition function which has the following form in Euclidean space:
\begin{equation}
      Z = \int [DB][D\Phi][D\Phi^{+}]\,e^{-S},     \lb{8y}
\end{equation}
where the action $S = \int d^4x L(x) + S_{gf}$ contains the Lagrangian
(\ref{5y}) written in Euclidean space and gauge fixing action $S_{gf}$.
Let us consider now a shift:
$ \Phi (x) = \Phi_b + {\hat \Phi}(x)$
with $\Phi_b$ as a background field and calculate the
following expression for the partition function in the one-loop
approximation:
$$
  Z = \int [DB][D\hat \Phi][D{\hat \Phi}^{+}]
   \exp\{ - S(B,\Phi_b)
   - \int d^4x [\frac{\delta S(\Phi)}{\delta \Phi(x)}|_{\Phi=
   \Phi_b}{\hat \Phi}(x) + h.c. ]\}\\
$$
\begin{equation}
    =\exp\{ - F(\Phi_b, g^2, \mu^2, \lambda)\}.      \lb{10y}
\end{equation}
Using the representation (\ref{7y}), we obtain the effective potential:
\begin{equation}
  V_{eff} = F(\phi_b, g^2, \mu^2, \lambda)        \lb{11y}
\end{equation}
given by the function $F$ of Eq.(\ref{10y}) for the constant background
field $ \Phi_b = \phi_b = \mbox{const}$. In this case the one--loop
effective potential for monopoles coincides with the expression of the
effective potential calculated by the authors of Ref.\ct{9} for scalar
electrodynamics and extended to the massive theory (see review \ct{10}).
As it was shown in Ref.\ct{9}, the effective potential
can be improved by consideration of the renormalization
group equation (RGE).

\section{Renormalization group equations in the Higgs monopole model}

The RGE for the effective potential means that the potential cannot
depend on a change in the arbitrary parameter --- renormalization scale $M$:
\begin{equation}
         \frac {dV_{eff}}{dM} = 0.             \lb{17y}
\end{equation}
The effects of changing it are absorbed into
changes in the coupling constants, masses and fields, giving so--called
running quantities.

Considering the RG improvement of the effective potential [8,9]
and choosing the evolution variable as
\begin{equation}
                   t = \log(\phi^2/M^2),     \lb{18y}
\end{equation}
we have the following RGE for the improved $V_{eff}(\phi^2)$
with $\phi^2\equiv \phi^2_b$ \ct{11}:
\begin{equation}
   (M^2\frac{\partial}{\partial M^2} + \beta_{\lambda}\frac{\partial}
{\partial\lambda} + \beta_g\frac{\partial}{\partial g^2} +
\beta_{(\mu^2)}\mu^2\frac{\partial}{\partial \mu^2} - \gamma\phi^2
\frac{\partial}{\partial \phi^2})V_{eff}(\phi^2) = 0,    \lb{19y}
\end{equation}
where $\gamma$ is the anomalous dimension and $\beta_{(\mu^2)}$,
$\beta_{\lambda}$ and $\beta_g$ are the RG $\beta$--functions for mass,
scalar and gauge couplings, respectively. RGE (\ref{19y}) leads to the
following form of the improved effective potential \ct{9}:
\begin{equation}
     V_{eff} = \frac{1}{2}\mu^2_{run}(t)G^2(t)\phi^2 +
                 \frac{1}{4}\lambda_{run}(t)G^4(t)\phi^4.  \lb{20y}
\end{equation}
In our case:
\begin{equation}
 G(t) = \exp[-\frac{1}{2}\int_0^t dt'\,\gamma\left(g_{run}(t'),
         \lambda_{run}(t')\right)].                         \lb{21y}
\end{equation}
A set of ordinary differential equations (RGE) corresponds to Eq.(\ref{19y}):
\begin{equation}
    \frac{d\lambda_{run}}{dt} = \beta_{\lambda}\left(g_{run}(t),\,
                    \lambda_{run}(t)\right),      \lb{22y}
\end{equation}
\begin{equation}
    \frac{d\mu^2_{run}}{dt} = \mu^2_{run}(t)\beta_{(\mu^2)}
             \left(g_{run}(t),\,\lambda_{run}(t)\right),
                                                  \lb{23y}
\end{equation}
\begin{equation}
    \frac{dg^2_{run}}{dt} = \beta_g\left(g_{run}(t),\,\lambda_{run}(t)\right).
                                           \lb{24y}
\end{equation}
So far as the mathematical structure of HMM is equivalent
to HSED, we can use all results of the scalar electrodynamics
in our calculations, replacing the electric charge $e$ and photon
field $A_{\mu}$ by magnetic charge $g$ and dual gauge field $B_{\mu}$.

The one--loop results for $\beta_{\lambda}^{(1)}$, $\beta_{\mu^2}^{(1)}$ and
$\gamma$ are given in
Ref.\ct{9} for scalar field with electric charge $e$, but it is easy to
rewrite them for monopoles with charge $g=g_{run}$:
\begin{equation}
\gamma^{(1)} = - \frac{3g_{run}^2}{16\pi^2},   \lb{30y}
\end{equation}
\begin{equation}
 \frac{d\lambda_{run}}{dt}\approx \beta_{\lambda}^{(1)} = \frac
1{16\pi^2} ( 3g^4_{run} +10 \lambda^2_{run} - 6\lambda_{run}g^2_{run}),
                                     \lb{31y}
\end{equation}
\begin{equation}
\frac{d\mu^2_{run}}{dt}\approx \beta_{(\mu^2)}^{(1)}
= \frac{\mu^2_{run}}{16\pi^2}( 4\lambda_{run} - 3g^2_{run} ),
                                                \lb{32y}
\end{equation}
\begin{equation}
    \frac{dg^2_{run}}{dt}\approx
     \beta_g^{(1)} = \frac{g^4_{run}}{48\pi^2}.  \lb{33y}
\end{equation}

The RG $\beta$--functions for different renormalizable gauge theories with
semisimple group have been calculated in the two--loop approximation
and even beyond. But in this paper we made use the
results of Refs.\ct{12} and \ct{13} for calculation of $\beta$--functions
and anomalous dimension in the two--loop approximation, applied to the
HMM with scalar monopole fields. The higher approximations essentially
depend on the renormalization scheme.
Thus, on the level of two--loop approximation we have for all
$\beta$--functions:
\begin{equation}
  \beta = \beta^{(1)} + \beta^{(2)},           \lb{34y}
\end{equation}
where
\begin{equation}
  \beta_{\lambda}^{(2)} = \frac{1}{(16\pi^2)^2}( - 25\lambda^3 +
   \frac{15}{2}g^2{\lambda}^2 - \frac{229}{12}g^4\lambda - \frac{59}{6}g^6),
                                                    \lb{35y}
\end{equation}
and
\begin{equation}
\beta_{(\mu^2)}^{(2)} = \frac{1}{(16\pi^2)^2}(\frac{31}{12}g^4 + 3\lambda^2).
                                             \lb{36y}
\end{equation}
The gauge coupling $\beta_g^{(2)}$--function is given by Ref.\ct{12}:
\begin{equation}
     \beta_g^{(2)} = \frac{g^6}{(16\pi^2)^2}.  \lb{37y}
\end{equation}
Anomalous dimension follows from calculations made in Ref.\ct{13}:
\begin{equation}
    \gamma^{(2)} = \frac{1}{(16\pi^2)^2}\frac{31}{12}g^4.
                                                   \lb{38y}
\end{equation}
In Eqs.(\ref{34y})--(\ref{38y}) and below, for simplicity, we have used the
following notations: $\lambda\equiv \lambda_{run}$, $g\equiv g_{run}$ and
$\mu\equiv \mu_{run}$.

\section{The phase diagram in the Higgs monopole model}

Now we  want to apply the effective potential calculation as a
technique for the getting phase diagram information for the condensation
of monopoles in HMM.
If the first local minimum occurs
at $\phi = 0$ and $V_{eff}(0) = 0$, it corresponds to the Coulomb--like phase.
In the case when the effective potential has the second local minimum at
$\phi = \phi_{min} \neq 0\,$ with $\,V_{eff}^{min}(\phi_{min}^2) < 0$,
we have the confinement phase. The phase transition between the
Coulomb--like and confinement phases is given by the condition when
the first local minimum at $\phi = 0$ is degenerate with the second minimum
at $\phi = \phi_0$.
These degenerate minima are shown in Fig.1 by the curve 1. They correspond
to the different vacua arising in this model. And the dashed curve 2
describes the appearance of two minima corresponding to the confinement
phases.

The conditions of the existence of degenerate vacua are given by the
following equations:
\begin{equation}
           V_{eff}(0) = V_{eff}(\phi_0^2) = 0,     \lb{39y}
\end{equation}
\begin{equation}
    \frac{\partial V_{eff}}{\partial \phi}|_{\phi=0} =
    \frac{\partial V_{eff}}{\partial \phi}|_{\phi=\phi_0} = 0,
\quad{\mbox{or}}\quad V'_{eff}(\phi_0^2)\equiv
\frac{\partial V_{eff}}{\partial \phi^2}|_{\phi=\phi_0} = 0,
                                                    \lb{40y}
\end{equation}
and inequalities
\begin{equation}
    \frac{\partial^2 V_{eff}}{\partial \phi^2}|_{\phi=0} > 0, \qquad
    \frac{\partial^2 V_{eff}}{\partial \phi^2}|_{\phi=\phi_0} > 0.
                                               \lb{41y}
\end{equation}
The first equation (\ref{39y}) applied to Eq.(\ref{20y}) gives:
\begin{equation}
    \mu^2_{run} = - \frac{1}{2} \lambda_{run}(t_0)\,\phi_0^2\, G^2(t_0),
\quad{\mbox{where}}\quad t_0 = \log(\phi_0^2/M^2).
                                    \lb{42y}
\end{equation}
It is easy to find the joint solution of equations
\begin{equation}
      V_{eff}(\phi_0^2) = V'_{eff}(\phi_0^2) = 0.       \lb{47y}
\end{equation}
Using RGE (\ref{22y}), (\ref{23y}) and Eqs.(\ref{40y})--(\ref{47y}),
we obtain:
\begin{equation}
 V'_{eff}(\phi_0^2) =\frac{1}{4}( - \lambda_{run}\beta_{(\mu^2)} +
\lambda_{run} + \beta_{\lambda} - \gamma \lambda_{run})G^4(t_0)\phi_0^2 = 0,
                                                    \lb{48y}
\end{equation}
or
\begin{equation}
    \beta_{\lambda} + \lambda_{run}(1 - \gamma - \beta_{(\mu^2)}) = 0.
                                            \lb{49y}
\end{equation}
Substituting in Eq.(\ref{49y}) the functions
$\beta_{\lambda}^{(1)},\,\beta_{(\mu^2)}^{(1)}$ and $\gamma^{(1)}$
given by Eqs.(\ref{30y})---(\ref{33y}), we obtain in the one--loop
approximation the following equation for the phase transition border:
\begin{equation}
     g^4_{PT} = - 2\lambda_{run}(\frac{8\pi^2}3 + \lambda_{run}).
                                                 \lb{50y}
\end{equation}
The curve (\ref{50y}) is represented on the phase diagram
$(\lambda_{run}; g^2_{run})$ of Fig.2 by the curve "1" which describes
the border between the Coulomb--like phase with $V_{eff} \ge 0$
and the confinement one with $V_{eff}^{min} < 0$. This border corresponds to
the one--loop approximation.

Using Eqs.(\ref{30y})-(\ref{38y}), we are able to construct
the phase transition border in the two--loop approximation.
Substituting these equations into Eq.(\ref{49y}), we obtain the following
phase transition border curve equation in the two--loop approximation:
\begin{equation}
 3y^2 - 16\pi^2 + 6x^2 + \frac{1}{16\pi^2}(28x^3 + \frac{15}{2}x^2y +
  \frac{97}{4}xy^2 - \frac{59}{6}y^3) = 0,            \lb{51y}
\end{equation}
where $x = - \lambda_{PT}$ and $y = g^2_{PT}$ are the phase transition
values of $ - \lambda_{run}$ and $g^2_{run}$.
Choosing the physical branch corresponding to $g^2 \ge 0$ and $g^2\to 0$,
when $\lambda \to 0$, we have received curve 2 on the phase diagram
$(\lambda_{run}; g^2_{run})$ shown in Fig.2. This curve
corresponds to the two--loop approximation and can be compared with
the  curve 1 of Fig.2, which describes the same phase transition border
calculated in the one--loop approximation.
It is easy to see that the accuracy of the 1--loop
approximation is not excellent and can commit errors of order 30\%.

According to the phase diagram drawn in Fig.2, the confinement phase
begins at $g^2 = g^2_{max}$ and exists under the phase transition border line
in the region $g^2 \le g^2_{max}$, where $e^2$ is large:
$e^2\ge (2\pi/g_{max})^2$ due to the Dirac relation:
\begin{equation}
   eg = 2\pi, \quad{\mbox{or}}\quad \alpha \tilde \alpha = \frac{1}{4}.
                                             \lb{54y}
\end{equation}
Therefore, we have:
$$
g^2_{crit} = g^2_{max1}\approx 18.61,
\quad
   \tilde \alpha_{crit} = \frac {g^2_{crit}}{4\pi}\approx 1.48,
\quad
  \alpha_{crit} = \frac{1}{4{\tilde \alpha}_{crit}}\approx 0.17\quad- \quad
{\mbox{in the one--loop approximation}},
$$
\begin{equation}
   g^2_{crit} = g^2_{max2}\approx
  15.11, \quad
   \tilde \alpha_{crit} = \frac {g^2_{crit}}{4\pi}\approx 1.20,\quad
        \alpha_{crit} = \frac{1}{4{\tilde \alpha}_{crit}}\approx 0.208
  \quad -\quad {\mbox{in the two--loop approximation}}.   \lb{55y}
\end{equation}
Comparing these results, we obtain the accuracy of
deviation between them of order 20\%.

The last result (\ref{55y}) coincides with the lattice values
obtained for the compact QED by Monte Carlo method \ct{7}:
\begin{equation}
{\alpha}_{crit}\approx{0.20\pm 0.015},\quad\quad
{\tilde \alpha}_{crit}\approx{1.25\pm 0.10}.      \lb{56}
\end{equation}
Writing Eq.(\ref{24y}) with $\beta_g$ function given by Eqs.(\ref{33y}),
(\ref{34y}), and (\ref{37y}), we have the following RGE for the monopole
charge in the two--loop approximation:
\begin{equation}
  \frac{dg^2_{run}}{dt}\approx \frac{g^4_{run}}{48\pi^2} +
    \frac{g^6_{run}}{(16\pi^2)^2},
\quad {\mbox{or}}\quad
   \frac{d\log{\tilde \alpha}}{dt}\approx \frac{\tilde \alpha}{12\pi}
            (1 + 3\frac{\tilde \alpha}{4\pi}).    \lb{56ay}
\end{equation}
The values (\ref{55y})  for $g^2_{crit} = g^2_{{max}1,2}$ indicate
that the contribution of two loops described by the second term of
Eq.(\ref{56ay}) is about 0.3, confirming the validity of perturbation theory.

In general, we are able to estimate the validity of two--loop approximation
for all $\beta$--functions and $\gamma$, calculating the corresponding
ratios of two--loop contributions to one--loop contributions
at the maxima of curves 1 and 2:
\begin{equation}
\begin{array}{|l|l|}
\hline %
&\\[-0.2cm]%
\lambda_{crit} = \lambda_{run}^{max1} \approx{-13.16}&\lambda_{crit} =
\lambda_{run}^{max2}\approx{-7.13}\\[0.5cm]
g^2_{crit} = g^2_{max1}\approx{18.61}& g^2_{crit} = g^2_{max2}
\approx{15.11}\\[0.5cm]
\frac{\ds\gamma^{(2)}}{\ds\gamma^{(1)}}\approx{-0.0080}&\frac{\ds
\gamma^{(2)}}{\ds\gamma^{(1)}}\approx{-0.0065}\\[0.5cm]
\frac{\ds\beta_{\mu^2}^{(2)}}{\ds\beta_{\mu^2}^{(1)}}\approx{-0.0826}
&\frac{\ds\beta_{\mu^2}^{(2)}}{\ds\beta_{\mu^2}^{(1)}}
\approx{-0.0637}\\[0.8cm]
\frac{\ds\beta_{\lambda}^{(2)}}{\ds\beta_{\lambda}^{(1)}}\approx{0.1564}
&\frac{\ds\beta_{\lambda}^{(2)}}
{\ds\beta_{\lambda}^{(1)}}\approx{0.0412}\\[0.8cm]
\frac{\ds\beta_g^{(2)}}{\ds\beta_g^{(1)}}\approx{0.3536}&\frac{\ds
\beta_g^{(2)}}{\ds\beta_g^{(1)}}\approx{0.2871}\\[0.5cm]
\hline
\end{array}
                                         \lb{57y}
\end{equation}
Here we see that all ratios are sufficiently small, i.e. all
two--loop contributions are small in comparison with one--loop contributions,
confirming the validity of perturbation theory in the 2--loop
approximation. The accuracy of deviation is worse
($\sim 30\%$) for $\beta_g$--function. But it is necessary to emphasize
that calculating the border curves 1 and 2 of Fig.2, we have not used
RGE (\ref{24y}) for monopole charge: $\beta_g$--function is absent in
Eq.(\ref{49y}). Therefore, the calculation of $g^2_{crit}$ according to
Eq.(\ref{51y}) does not depend on the approximation of $\beta_g$ function.
The above--mentioned $\beta_g$--function appears only in the second order
derivative of $V_{eff}$ which is related with the monopole mass $m$
(see Refs.[2,3]).

Eqs.(\ref{55y}) give the following result:
\begin{equation}
 \alpha_{crit}^{-1}\approx 5,      \lb{56y}
\end{equation}
which is important for the phase transition at the Planck scale
predicted by the Multiple Point Model (MPM).

\section{Multiple Point Model and Critical Values
of the U(1) and SU(N) Fine Structure Constants}

Investigating the phase transition in HMM,
we had pursued two objects: from one side, we had an aim to
explain the lattice results, from the other side, we were interested
in the predictions of MPM.

\subsection{Anti-grand unification theory}

Most efforts to explain the Standard Model (SM) describing well all
experimental results known today are devoted to Grand Unification
Theories (GUTs). The supersymmetric extension of the SM consists of taking the
SM and adding the corresponding supersymmetric partners.
Unfortunately, at present time experiment does not indicate any manifestation
of the supersymmetry. In this connection, the Anti--Grand Unification
Theory (AGUT) was developed in Refs.[13-17, 4] as a realistic
alternative to SUSY GUTs.  According to this theory, supersymmetry does not
come into the existence up to the Planck energy scale:
$M_{Pl}=1.22\cdot 10^{19}$ GeV.

The Standard Model (SM) is based on the group SMG:
\begin{equation}
SMG = SU(3)_c\times SU(2)_L\times U(1)_Y.  \lb{2}
\end{equation}
AGUT suggests that at the energy scale $\mu_G\sim \mu_{Pl}=M_{Pl}$ there
exists the more fundamental group $G$ containing $N_{gen}$ copies of the
Standard Model Group SMG:  \begin{equation} G = SMG_1\times SMG_2\times...\times
SMG_{N_{gen}}\equiv (SMG)^{N_{gen}}, \lb{76y} \end{equation} where $N_{gen}$ designates
the number of quark and lepton generations (families).

If $N_{gen}=3$ (as AGUT predicts), then the fundamental gauge group G is:
\begin{equation}
    G = (SMG)^3 = SMG_{1st\, gen.}\times SMG_{2nd\, gen.}\times SMG_{3rd\,
                                       gen.},  \lb{77y}
\end{equation}
or the generalized ones:
\begin{equation}
         G_f = (SMG)^3\times U(1)_f
         \quad\ct{17}, \quad {\mbox{or}}\quad
      G_{\mbox{ext}} = (SMG\times U(1)_{B-L})^3 \quad\ct{19},      \lb{78y}
\end{equation}
which were suggested by the fitting of fermion masses of the SM
(see Refs.\ct{17}), or by the see--saw mechanism with right-handed
neutrinos \ct{19}.

\subsection{Multiple Point Principle}

AGUT approach is used in conjuction with the Multiple Point
Principle proposed in Ref.\ct{4}.
According to this principle Nature seeks a special point --- the Multiple
Critical Point (MCP) --- which is a point on the phase diagram of the
fundamental regulirized gauge theory G (or $G_f$, or $G_{ext}$), where
the vacua of all fields existing in Nature are degenerate having the same
vacuum energy density.
Such a phase diagram has axes given by all coupling constants
considered in theory. Then all (or just many) numbers of phases
meet at the MCP.
MPM assumes the existence of MCP at the Planck scale,
insofar as gravity may be "critical" at the Planck scale.

The philosophy of MPM leads to the necessity
to investigate the phase transition in different gauge theories.
A lattice model of gauge theories is the most convenient formalism for the
realization of the MPM ideas. As it was mentioned above,
in the simplest case we can imagine our
space--time as a regular hypercubic (3+1)--lattice with the parameter $a$
equal to the fundamental (Planck) scale: $a = \lambda_P = 1/M_{Pl}$.

\subsection{AGUT-MPM prediction of the Planck scale values of the
U(1), SU(2) and SU(3) fine structure constants}

The usual definition of the SM coupling constants:
\begin{equation}
  \alpha_1 = \frac{5}{3}\frac{\alpha}{\cos^2\theta_{\ov{MS}}},\quad
  \alpha_2 = \frac{\alpha}{\sin^2\theta_{\ov{MS}}},\quad
  \alpha_3 \equiv \alpha_s = \frac {g^2_s}{4\pi},     \lb{81y}
\end{equation}
where $\alpha$ and $\alpha_s$ are the electromagnetic and strong
fine structure constants, respectively, is given in the Modified
minimal subtraction scheme ($\ov{MS}$).
Here $\theta_{\ov{MS}}$ is the Weinberg weak angle in $\ov{MS}$ scheme.
Using RGE with experimentally
established parameters, it is possible to extrapolate the experimental
values of three inverse running constants $\alpha_i^{-1}(\mu)$
(here $\mu$ is an energy scale and i=1,2,3 correspond to U(1),
SU(2) and SU(3) groups of the SM) from the Electroweak scale to the Planck
scale. The precision of the LEP data allows to make this extrapolation
with small errors (see \ct{20}). Assuming that these RGEs for
$\alpha_i^{-1}(\mu)$ contain only the contributions of the SM particles
up to $\mu\approx \mu_{Pl}$ and doing the extrapolation with one
Higgs doublet under the assumption of a "desert", the following results
for the inverses $\alpha_{Y,2,3}^{-1}$ (here $\alpha_Y\equiv \frac{3}{5}
\alpha_1$) were obtained in Ref.\ct{4} (compare with \ct{20}):
\begin{equation}
   \alpha_Y^{-1}(\mu_{Pl})\approx 55.5; \quad
   \alpha_2^{-1}(\mu_{Pl})\approx 49.5; \quad
   \alpha_3^{-1}(\mu_{Pl})\approx 54.0.
                                                        \lb{82y}
\end{equation}
The extrapolation of $\alpha_{Y,2,3}^{-1}(\mu)$ up to the point
$\mu=\mu_{Pl}$ is shown in Fig.3.

According to AGUT, at some point $\mu=\mu_G < \mu_{Pl}$ (but near
$\mu_{Pl}$) the fundamental group $G$ (or $G_f$, or $G_{\mbox{ext}}$)
undergoes spontaneous breakdown to its diagonal subgroup:
\begin{equation}
      G \longrightarrow G_{diag.subgr.} = \{g,g,g || g\in SMG\},
                                                          \lb{83y}
\end{equation}
which is identified with the usual (low--energy) group SMG.

The AGUT prediction of the values of $\alpha_i(\mu)$ at $\mu=\mu_{Pl}$
is based on the MPM assumptions, and gives these values
in terms of the corresponding critical couplings $\alpha_{i,crit} $
[13-15,4]:
\begin{equation}
            \alpha_i(\mu_{Pl}) = \frac {\alpha_{i,crit}}{N_{gen}}
                       = \frac{\alpha_{i,crit}}{3}
                \quad{\mbox{for}}\quad i=2,3,       \lb{84y}
\end{equation}
and
\begin{equation}
\alpha_1(\mu_{Pl}) = \frac{\alpha_{1,crit}}{\frac{1}{2}N_{gen}(N_{gen} + 1)}
                   = \frac{\alpha_{1,crit}}{6} \quad{\mbox{for}}\quad U(1).
                                      \lb{85y}
\end{equation}
It was assumed in Ref.\ct{4} that the MCP values
$\alpha_{i,crit}$ in Eqs.(\ref{84y}) and (\ref{85y}) coincide with
the triple point values of the effective fine structure
constants given by the lattice SU(3)--, SU(2)-- and U(1)--gauge theories.

If the point $\mu=\mu_G$ is very close to the Planck scale
$\mu=\mu_{Pl}$, then according to Eqs.(\ref{82y}) and (\ref{85y}), we have:
\begin{equation}
         \alpha_{1st\, gen.}^{-1}\approx
    \alpha_{2nd\, gen.}^{-1}\approx \alpha_{3rd\, gen.}^{-1}\approx
    \frac{\alpha_Y^{-1}(\mu_G)}{6}\approx 9,        \lb{88y}
\end{equation}
what is almost equal to the value:
\begin{equation}
            \alpha_{crit.,theor}^{-1}\approx 8      \lb{89y}
\end{equation}
obtained theoretically by Parisi improvement method
for the Coulomb-like phase [4,6]. The critical value (\ref{89y})
is close to the lattice and HMM ones: see Eq.(\ref{56y}).
This means that in the U(1) sector of AGUT we have $\alpha $ near
the critical point, and we can expect the existence of MCP
at the Planck scale.

\vspace*{30mm}

\bc
\noindent\includegraphics[width=145mm, height=130mm]{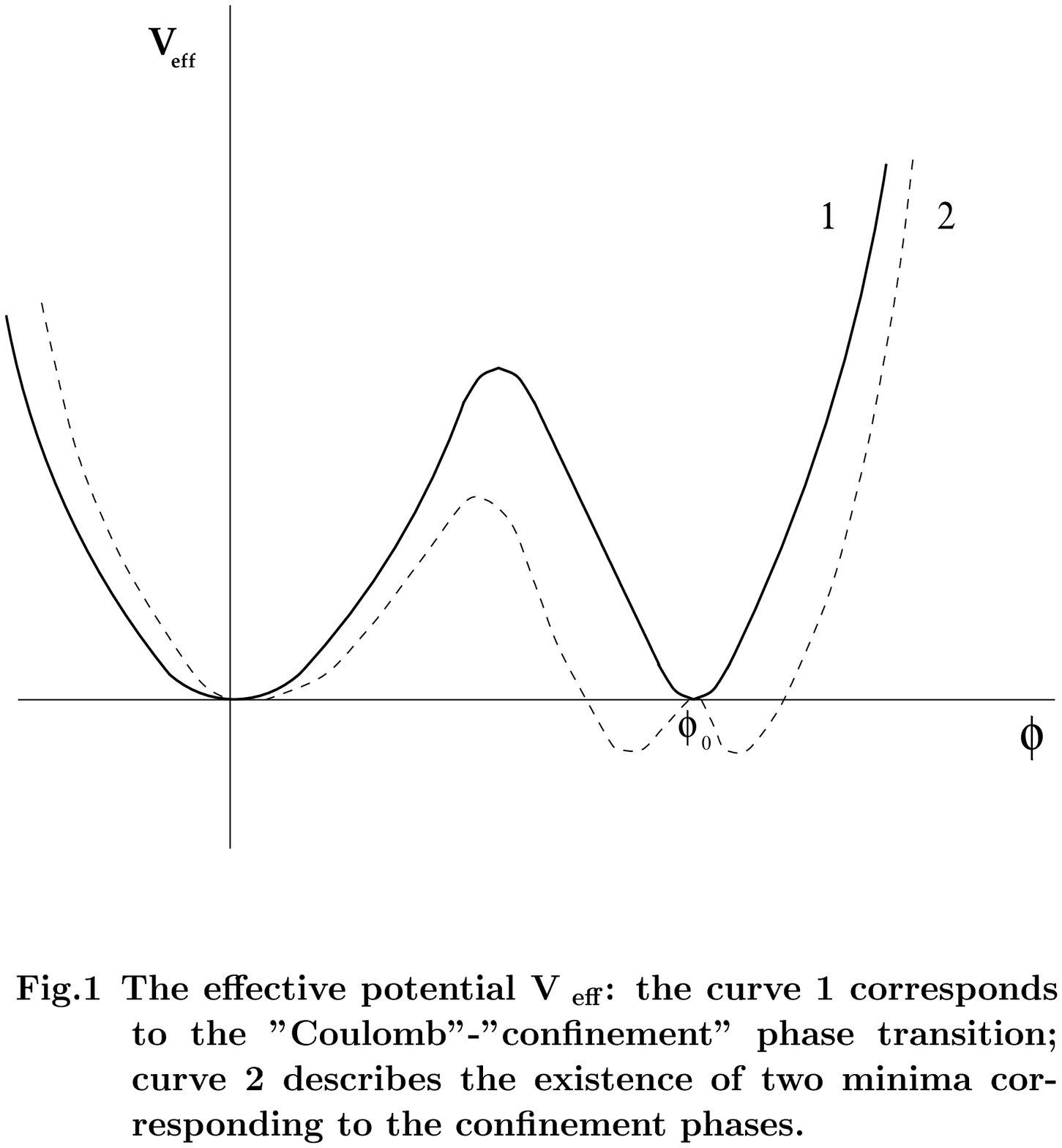}
\ec

\bc
\noindent\includegraphics[width=120mm,height=110mm]{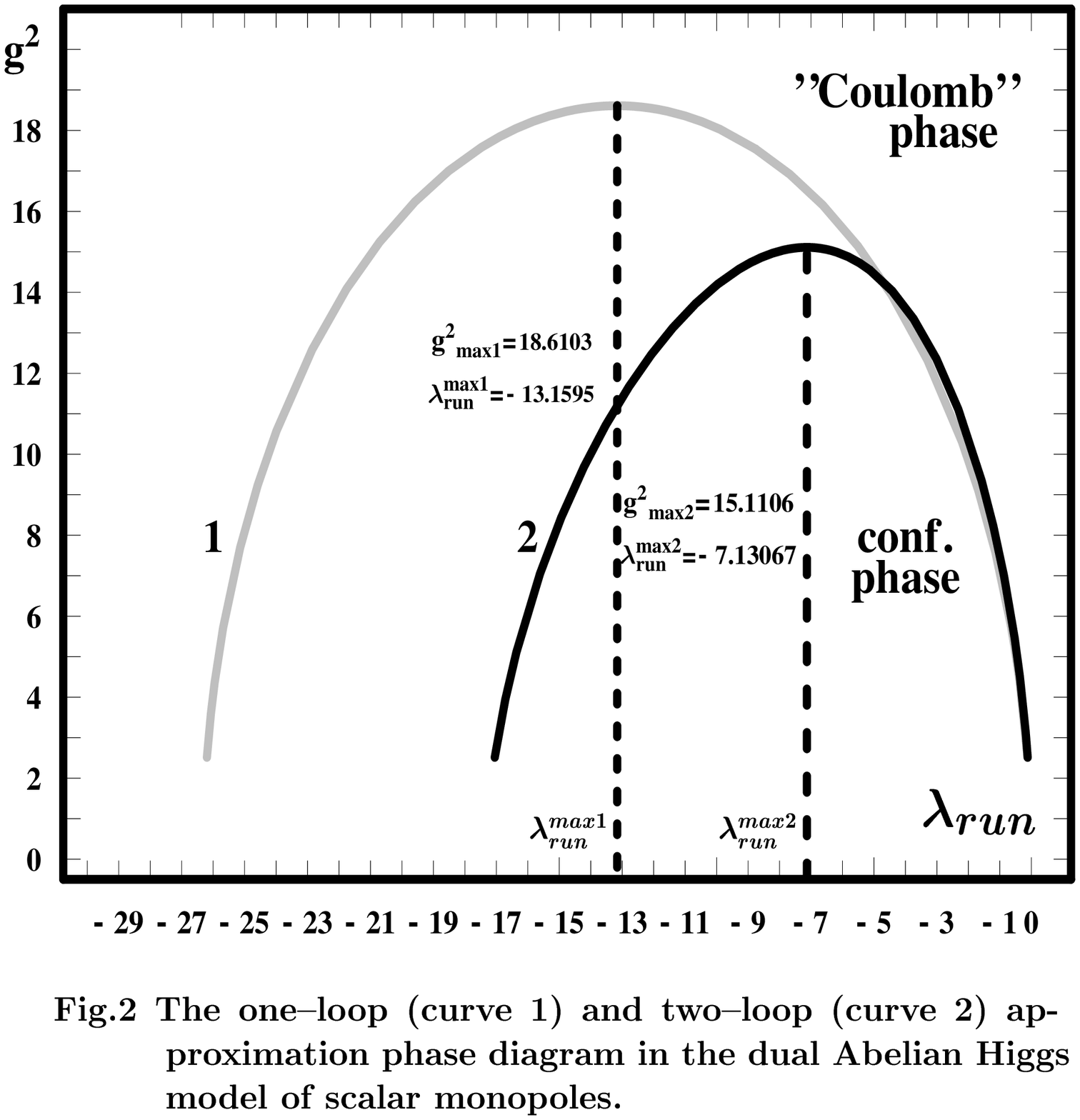}
\ec

\bc
\noindent\includegraphics[width=120mm, height=110mm]{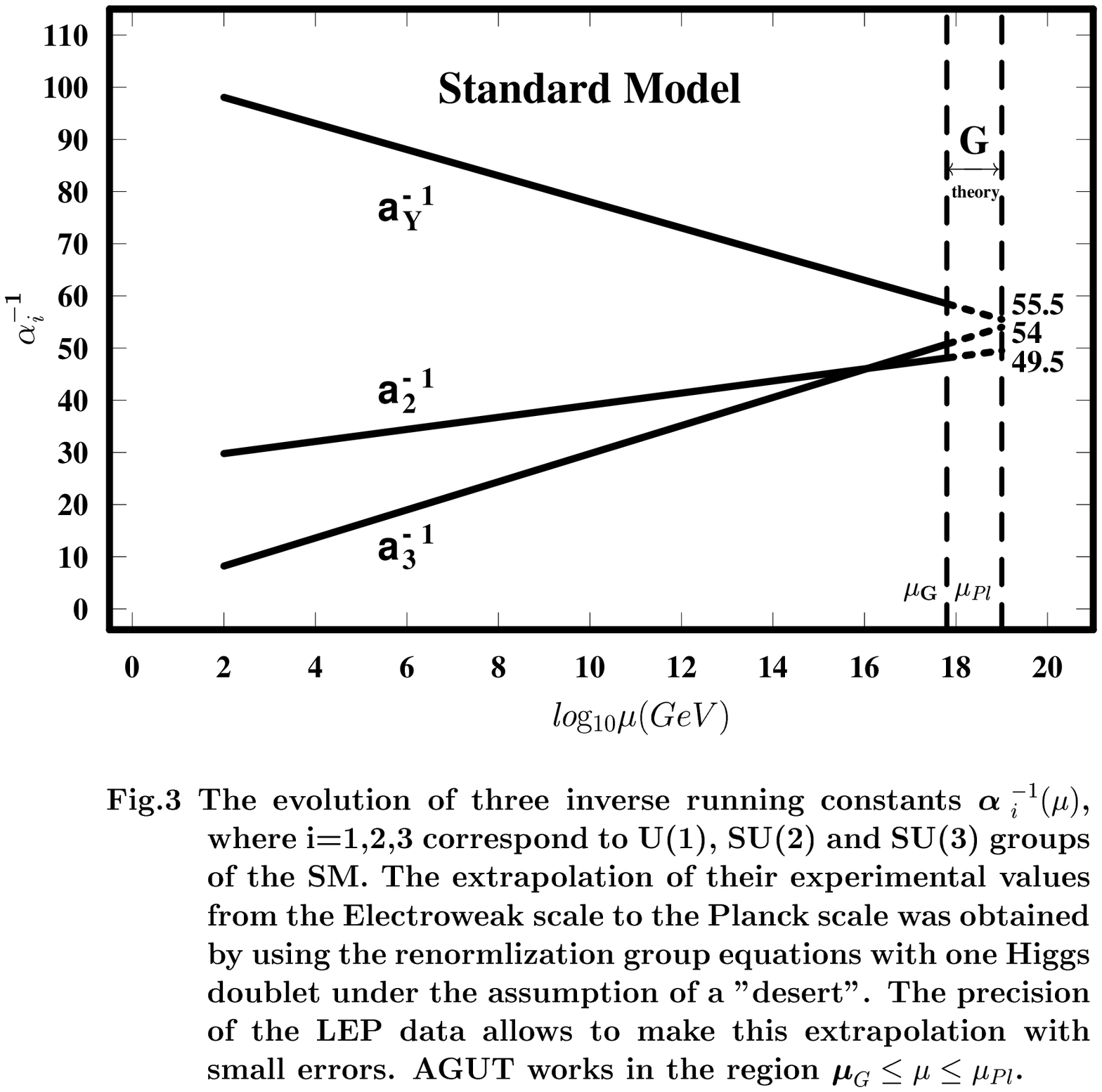}
\ec


\end{document}